\documentclass[superscriptaddress,aps,prl,twocolumn]{revtex4}
\def\beq{\begin{equation}}
\def\eeq{\end{equation}}
\def\bea{\begin{eqnarray}}
\def\eea{\end{eqnarray}}
\def\lb{\langle}
\def\rb{\rangle}
\def\be{\begin{equation}}
\def\ee{\end{equation}}
\usepackage{graphicx}

\begin{document}

\title{Unitary Fermi Gas in a Harmonic Trap}
\stepcounter{mpfootnote}
\author {S. Y. Chang and G. F. Bertsch}
\affiliation { Department of Physics and Institute for Nuclear Theory,
             Box 351560, University of Washington, Seattle, WA 98195}
\date{\today}

\begin{abstract}
We present an {\it ab initio} calculation of small numbers of trapped,
strongly interacting fermions using the Green's Function Monte Carlo method (GFMC). 
The ground state energy, density profile and pairing gap are calculated
for particle numbers $N = 2 \sim 22$ using the parameter-free ``unitary"
interaction.  Trial wave functions are taken of the form of correlated pairs
in a harmonic oscillator basis. We find that the lowest energies are 
obtained with a minimum explicit
pair correlation beyond that needed to exploit the degeneracy of oscillator
states.  We find that energies can be well fitted by the expression
$a_{TF} E_{TF} + \Delta\, {\rm mod}(N,2)$ where $E_{TF}$ is the Thomas-Fermi
energy of a noninteracting gas in the trap and $\Delta$ is a pairing
gap.  There is no evidence of a shell correction energy in the
systematics, but the density distributions show pronounced shell effects.
We find the value $\Delta= 0.7\pm 0.2\omega$ for the pairing gap.  This is
smaller than the value found for the uniform gas at a density corresponding
to the central density of the trapped gas.
\end{abstract}
\maketitle

The physics of cold trapped atoms in quantum condensates has seen 
remarkable advances on the experimental front,  
particularly with the possibility to study pairing condensates in fermionic
systems\cite{ohara02,barten04,bourdel04,chin04,zwier05,kinast05,stewart06}.  
Many features of systems in the size range 
$N\sim 10^4-10^6$ are now well-explored, but the small-$N$
limit is also of great interest for optical lattices.  In this work we investigate
the properties of small systems of trapped fermionic atoms
using the Green's Function Monte Carlo technique (GFMC) that has
been successful in the study of the homogeneous
gas\cite{ca03}.  The small systems are in some ways 
more challenging because simplifications that follow from
translational invariance are not present.
Our main goal here is to see how the bulk behavior evolves as a
function of the number of atoms and to provide benchmark
{\it ab initio} results to test other theoretical methods.
 The Hamiltonian for interacting atoms in a spherical harmonic trap
is given by
\beq
{\cal H} =  \sum\limits_{i=1}^N \left[ \frac{1}{2m }p_i^2 + 
\frac{1}{2}m\omega^2 r_i^2 \right] + \sum\limits_{i=1}^{N_\uparrow}\sum\limits_{j=1}^{N_\downarrow}
v(r_{ij})
\eeq
where $\omega$ is the trap frequency and $v(r)$ is the  
interaction between the atoms of opposite spin states.
 We will use units with $\hbar=1$.
 The interaction is chosen to approach the so-called unitary limit, meaning
that it supports a two-body bound state at zero
energy as well as having a range much shorter than any other length
scales in the Hamiltonian.  For technical reasons, we keep interaction range finite
in the GFMC using the form
\beq
v(r) = -{ 8\over m r_0^2 \cosh^2(2 r/r_0)}.
\eeq
The effective range of the potential is $r_0$; the short-range limit
is $r_0\sqrt{m\omega} \ll 1$.
The results are presented in the following sections, together
with some comparison to the expectations based on the local
density approximation (LDA).

To apply the GFMC, one starts with a trial wave function $\Psi_T$
that is antisymmetrized according to the fermion
statistics of the particles.  The GFMC gives the lowest energy state
in the space of wave functions that have the same
nodal structure as $\Psi_T$. 
We tried several approaches to parameterize $\Psi_T$.  For even
particle number $N$, they can all
be expressed in the form
\beq
\Psi_T = \left( \prod\limits_{i,j} f_{ij}\right){\cal A} \left[
 \phi^{(2)}({\bf r}_1, {\bf r}_{N_\uparrow+1}) \cdots \phi^{(2)}({\bf r}_{N_\uparrow}, {\bf r}_{N_\uparrow+N_\downarrow}) \right].
\label{eq_fst}
\eeq
Here $\phi^{(2)}(i,j)$ is a pair wave function,
$N_\uparrow=N_\downarrow=N/2$, and  $\Pi f$ is a Jastrow
correlation factor. The antisymetrization operation $\cal{A} $ is carried out by
evaluating determinant of an $N/2$-dimensional matrix.
For systems with an odd number of particles, we need to include an
unpaired particle in the wave function.  We define an orbital
wave function $\psi({\bf r})$ for the extra particle and it is included 
by adding an extra row and column to the determinant \cite[Eq. 14]{ca03}
in the antisymmetrization.
We have mostly investigated trial wave functions where the pair state
takes the form \cite{he02a,he02b},
\bea
\phi^{(2)}({\bf r}_1, {\bf r}_2) & = &
\sum_{\Lambda=0}^{\Lambda_c} \alpha_{\Lambda} \sum_l \sum_{n'\le n} \sum\limits_{m} 
(-1)^{l+m}/\sqrt{2l+1} \times \nonumber \\
 & & \psi_{nlm}({\bf r}_1)\psi^*_{n'l-m}({\bf r}_2). 
\label{eq_pairing}
\eea
Here $\psi_{nlm}$ is the oscillator state  
labeled by radial quantum number $n$ and angular momentum quantum numbers
$l,m$, the oscillator shell is
$\Lambda=2 n + l$, and $\Lambda_c$ is a shell cutoff. 
Clebsch-Gordan coefficients $(-1)^{l+m}/\sqrt{2l+1}$ allows that
the the pairs with angular momenta $(l,m)$ and $(l,-m)$ to form
zero total angular momentum state. The Ansatz Eq.(\ref{eq_pairing}) is analogous to the pair
wave function used to calculation the uniform system
\cite{ca03}. There the particle orbitals were plane waves and each was 
paired with the orbital of opposite momentum. This pair state allows for
intra-shell ($n=n'$) as well as multi-shell ($n \ne n'$) pairings. 
At shell closures such as $N=8,20$ the trial function is equivalent to
the Slater determinant of harmonic oscillator orbitals when the cutoff
is at the highest occupied shell and multi-shell pairings are neglected.
 We have also considered taking the pair wave function as the eigenstate of the two-particle Hamiltonian,
requiring in principle an infinite cutoff in the oscillator representation.
We call this case as 2B.

We carry out the GFMC in the usual way described in Ref. \cite{kalos74}.
The ground state wave function is projected out of the trial
wave function by evolving it in imaginary time, and the energy
is taken by the normalized matrix element of the Hamiltonian
operator.  This may be expressed as 
\beq
| \Psi_0\rb = \lim_{\tau \rightarrow \infty}e^{-{\cal{H}} \tau} |\Psi_T \rb
\eeq
and $ E_0 = \lim_{\tau \rightarrow \infty}  {\lb \Psi_T {\cal H} 
e^{-{\cal{H}} \tau} |\Psi_T\rb / \lb \Psi_T 
e^{-{\cal{H}} \tau} |\Psi_T\rb}$.
The integral is evaluated by the Monte Carlo method, carrying out the
exponentiation by the expansion
$e^{-{\cal{H}} \tau}\approx (e^{-V \Delta\tau/2} e^{-T \Delta \tau} e^{-V \Delta\tau/2})^M$ and using path sampling.
Our target accuracy is 1\% on the energies.  This is achieved by 
taking numerical parameters 
$\Delta \tau = 0.04 \omega  $ and $15,000< M < 30,000$.  
In practice, the convergence to the ground state is reached in the first few
thousands of steps. The Monte Carlo sample points that leave the
region where $|\Psi_T \rb > 0$ are discarded.  This nodal constraint avoids the
signal decay known as `fermion sign problem'.
The energies depend on the range parameter $r_0$ only in the 
combination $r_0 \sqrt{ m \omega}$ which we set to 0.1.  We believe
this is small enough to give energies that approach the contact limit to
within 1\%.  Smaller values of range parameter are possible but 
increase the statistical fluctuations of the Monte Carlo integration.

The cases $N=2,3$ are special in that analytic solutions are known.  For the
$N=2$ system, the Jastrow correlation factor can be defined to give the 
exact wave function and energy $E = 2 \omega$.  The $N=3$ system gives the first real test of the theory. 
The exact energy is $E=4.27...\omega$, given by the solution of a
transcendental equation in one variable\cite{we06}.
Using Eq. \ref{eq_pairing} with a single
term ($\Lambda_c = 0$) we find an energy of $4.28\pm0.04\omega$ in 
close agreement with the exact value.  In contrast, taking the pair wave function as 
the two-particle eigenstate gave a significant difference, $4.41\pm0.02\omega$.

We now turn to the larger systems and determine the parameters
in Eq. \ref{eq_pairing}.  As mentioned earlier, at the shell closures taking the
cutoff at the highest occupied shell gives the harmonic oscillator
Slater determinant (HOSD).  We also tried Slater determinants for
mid-shell systems, but typically they break rotational symmetry
and give a significantly higher energy.  Use of Eq. \ref{eq_pairing}
guarantees that $\Psi_T$ will be rotational invariant (for even $N$).  One
parameterization we explored was to use the results of Ref. \cite{ca03}
to guide the choice of $\Lambda_c$ and the $\alpha_\Lambda$.  This
gives a rather open trial wave function, having significant particle
excitation out of the lower shells.  Another
choice is to take each $\alpha_\Lambda$
proportional to the shell occupancy of the HOSD, which we call I1.  
For mid-shell systems, this also produces significant excitations out of the
nominally closed shells.  Both these schemes gave poorer energies
than we could achieve by taking amplitudes $\alpha_\Lambda$ that
maximize the occupancy of the filled shells of the HOSD, and 
have no occupancy in the nominally empty shells. 
Values of $\alpha_\Lambda$ that approach this compact limit (CL)
are given in Table \ref{tab_params}. As seen in this table, $\alpha_\Lambda$
parameters are not very sensitive in the ranges limited by the shell closures and
are kept constant.

For odd $N$, we take the pair wave function the same as in the
neighboring even system, and the orbital of the odd particle as
an oscillator state of the filling shell.  Thus, in the range
$N=3-7$ the orbital is a $p$-shell orbital with $(n,l)=(0,1)$.
Starting at $\Lambda=2$ there is a choice of orbitals, eg.
$(n,l) = (0,2)$ or $(1,0)$ for $N=9$.  We found for the
$N=9,11,13,15$ and $19$ systems, the energies are 
degenerate within the statistical errors and it was not possible to 
determine the density preference of the excitation.
For these cases, we simply took the odd
orbital to be one with the highest value of $l$.

\begin{figure}
\includegraphics[angle=-90,width= \columnwidth]{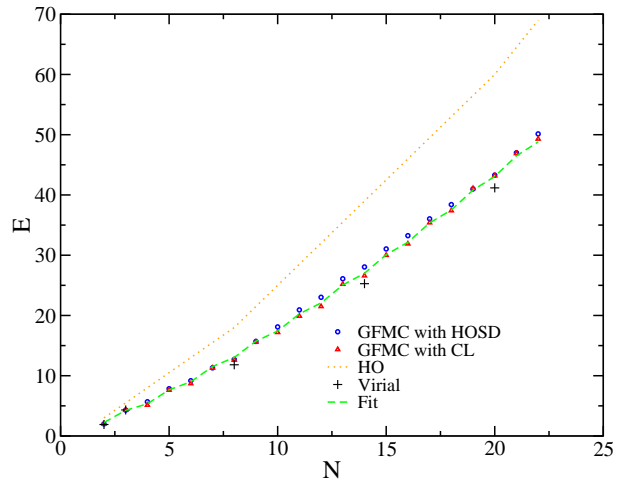}
\caption{(color online) Energy systematics of the trapped unitary
Fermi Gas.  Circles, GFMC calculated with the HOSD trial function;
triangles, GFMC with the CL trial function; crosses, virial formula.
The dotted line is the energy of the HOSD for free particles.  The dashed
line is the fit (Eq. \ref{eq_fit}) to the CL calculated energies.
The unit of energies is $\omega $.}
\label{fig_results}
\end{figure}

The calculated energies are summarized in Table \ref{tab_nrgs} for the
paired wave function in the compact limit.  The
statistical errors of the GFMC are given in parenthesis.  One
sees that an accuracy of 1\% is achieved with the numerical procedure
we described earlier.  In Fig. \ref{fig_results} we show a plot of the energies including
the results from the HOSD trial wave function.  As expected, the
energies are the same at the shell closures but the HOSD gives higher
energies in mid-shell.  As an example of the sensitivity of the energy
to the detailed assumption about the pair state, the results for 
$N=12$ are: $E_{CL}= 21.5(3)$, $E_{2B}=22.3(2)$ and $E_{I1} = 22.4(3)$.  One sees
that the energies are actually rather close.  However, $E_{CL}$ is consistently
$2-4\%$ below other pairing node assumptions in the range of $N$ considered. 
In case of the simple HOSD, $E_{HOSD} = 23.0(1)$ which is $~7\%$ above the CL energy.
This gives some confidence that the assumed nodal structure of $\Psi_T$ is adequate for
our purposes.  We will comment on the nodal structure again later.

\begin{table}
\caption{Parametrization of $\alpha_\Lambda$ (CL).}
\begin{tabular}{cccc|c}
$\alpha_0$ & $\alpha_1$ & $\alpha_2$ & $\alpha_3$ & \\
\hline
1.0 & 0 & 0 & 0 & N = 2,3 \\
1.0 & 0.1 & 0  & 0  & 4 $\le$ N $\le$ 7 \\
1.0 & 1.0 & 0 & 0 & N = 8,9\\
1.0 & 0.5 & 0.01 & 0  & 10 $\le$ N $\le$ 19 \\
1.0 & 1.0 & 1.0 & 0 & N = 20,21\\
1.0 & 1.0 & 0.5  & 0.01 &  N $\ge$ 22 \\
\hline
\end{tabular}
\label{tab_params}
\end{table}

\begin{table}
\caption{GFMC energies for the unitary trapped fermion
gas with the CL pair functions. Also shown are the energies 
of noninteracting gas (HOSD). 
The unit of energy is  $\omega$.}
\begin{tabular}{|c|cc||c|cc|}
\hline
N  & HOSD & GFMC   &   N & HOSD & GFMC  \\
\hline
2  & 3  & 2.01(2)    & 13 &35.5 &25.2(3)    \\
3  & 5.5 & 4.28(4)   & 14 &39 & 26.6(4)    \\
4  & 8    &5.1(1)    & 15 &42.5 &30.0(1) \\
5  & 10.5 &7.6(1)    & 16 &46 &31.9(3)  \\
6  & 13& 8.7(1)      & 17 &49.5 &35.4(2)   \\
7  & 15.5 &11.3(1)   & 18 &53&37.4(3)  \\
8  & 18 &12.6(1)     & 19 &56.5 &41.1(3)  \\
9  & 21.5 &15.6(2)   & 20 &60 &43.2(4)   \\
10 & 25 &17.2(2)     & 21 & 64.5& 46.9(2)\\
11 & 28.5& 19.9(2)   & 22 & 69 & 49.3(1)\\
12 & 32 & 21.5(3)    &    &  &  \\
\hline
\end{tabular}
\label{tab_nrgs}
\end{table}

These results show that the pairing is less important in the
trial wave function for the finite systems than it is in
the uniform gas.  While in the homogeneous gas a BCS treatment of the
trial function lowers the energy by more
than 20$\%$ ($E_{SF}=0.42 E_{FG}$ and $E_{normal} = 0.56 E_{FG}$),
the difference in energy between the both phases of the trapped gas does
not exceed $7\%$ at the most in the open shell configuration $N= 12$.
At the shell closures, the BCS treatment does not offer any improvement

There is a virial theorem for unitary trapped gases given by\cite{castin06}
$ E_0 =  2N \lb {\cal U} \rb$, where $ \lb {\cal U} \rb =  
\frac{1}{2} m \omega^2 \lb r^2\rb $.  The theory can thus be
tested by independently calculating the expectation value of the trapping potential.
Expectation values of operators are usually estimated in the GFMC by the
expression
$\lb \Psi_0|{\cal U}|\Psi_0 \rb
\approx 2 \lb {\cal U} \rb_{GFMC}- \lb{\cal U}\rb_{var.} $. 
Using this estimate of  $\lb {\cal U}\rb$, we find energies somewhat lower than obtained by direct GFMC calculation (see Fig. \ref{fig_results}). This could be due to the errors associated with the extrapolation formula for expectation values.  

We now examine how well the energies fit the asymptotic theory
for large nonuniform systems.  The first term in the theory is 
the Thomas-Fermi (TF)  approximation \cite{papen05}; the TF approximation to
the trapped unitary Fermi gas is $ E_{TF}(\xi) = \xi^{1/2} \omega (3 N)^{4/3}/4$,
where  $\xi$ is the universal
constant relating the energy of the uniform gas to that
of the free Fermi gas. 
Adding a second term in the expansion gives a better description of the energy of 
the harmonic oscillator energy of the trapped gas in the large $N$ limit \cite{br97}.
We therefore will include that in the fit to the 
energies, using the form 
\beq
E'_{TF}(\xi) = \xi^{1/2} \omega \left(\frac{(3 N)^{4/3}}{4} + \frac{(3N)^{2/3}}{8}\right).
\label{eq_fit}
\eeq 
As it may be seen from Fig. \ref{fig_results}, there is also
a significant odd-even variation in the energies.  We shall
include this effect as well by fitting to the function
\be
E= E'_{TF}(\xi) + \Delta {\rm mod} (N,2)
\ee
The result of the fit is shown by the dashed line in Fig. \ref{fig_results}.
The fit value of $\xi$ is $\xi= 0.50$. This is somewhat higher
than the bulk value $0.42-0.44$.  This suggests that the convergence to 
the bulk is rather slow.  One might expect to see shell effects in 
the energies once the smooth trends have been taken out.  The HO
energies, for example, oscillate around $E_{TF}(0)$ with (negative)
peaks at the shell closures.  The effect is visible in the abrupt
change of slop of the HO curve in Fig. \ref{fig_results}.  However, in our fit
to the calculated energies, we do not see a visible shell closure
effect.  In the fit we find for the parameter $\Delta$ the value
$\Delta= 0.6 \omega$, in accordance with the average
of the odd-even staggering of the energy $\Delta = 0.7(2) \omega$.
If the pairing gap were controlled by the density at the center,
it would be much larger; of the order of shell spacing or
higher.  On the other hand, if the odd particle is most
sensitive to the surface region, the pair effect could be
smaller. A more systematic approach to LDA exists where the
superfluid correlation is introduced $\it ab~initio$ \cite{bulgac02a,bulgac02b}.

\begin{figure}
\includegraphics[angle=-90,width= \columnwidth]{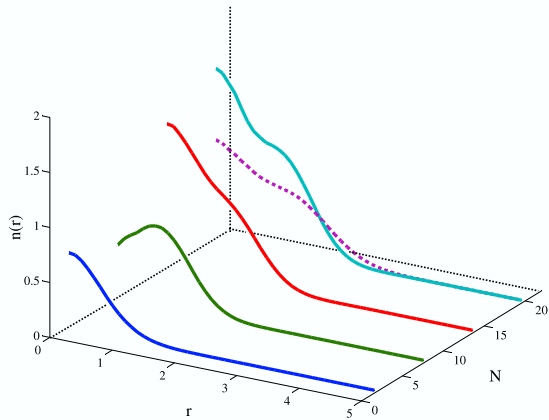}
\caption{(color online) Radial densities for $N=2,8,14$ and $20$. For $N=20$, free particle density distribution is also shown(dotted line). The unit
on the radial axis is $\frac{1}{\sqrt{m \omega}}$.}
\label{fig_densities}
\end{figure}

We now turn to the density distributions calculated with the GFMC.
We determine the density by binning the values obtained by
the Monte Carlo sampling.  With $\sim 15000$ paths and 1000 samples per
path there is adequate statistics to get details of the density
distribution well beyond the mean square radii. Fig. \ref{fig_densities} shows
the calculated densities for $N=2,8,14,$ and $20$.  We notice that
the central densities show a pronounced dip for $N=8$ and
a peak for $N=20$.  These are characteristic of shell structure,
depending on whether the highest occupied shell has $s$-wave
orbitals or not.  Fig. \ref{fig_densities} also shows the HO density for $N=20$.
The density of the interacting system is more compact, as required
by its lower energy and the virial theorem.  The central peak has
roughly the same relative shape in the two cases.  Thus the
basic HO pattern is maintained, even though the system shrinks
in size.

Let us return again to the problem of the trial function and its fixed
nodal structure.  It is interesting to ask how different are the 
nodal positions for the different $\Psi_T$'s.  We can characterize the trial
wave functions by their relative overlaps of the sign domains. If we define 
$$x = \frac{1}{N_S} \sum\limits_{i=1}^{N_S} \frac{1}{2} \left(1 +
 \frac{\Psi^*_{T1}\Psi_{T2} }{|\Psi^*_{T1} \Psi_{T2}|} \right)\times
100\%,$$
the nodal overlap between $\Psi_{T1} $ and $\Psi_{T2} $ is given as
$\mbox{\it Nodal overlap} = \max[100\%-x,x]$.
From this definition, nodal overlap ranges $50\% \sim 100\%$. Because of the strong
suppression of the superfluidity at the shell closures the nodal
overlap between the normal and the superfluid node wave functions
seems to be the largest $\sim 100\%$ at the shell closures and has low values $\sim 85\%$ at $N=5$ and $\sim 67\%$ at $N=14$.

We believe our computed energy systematics is reliable enough
to arrive at the following conclusions. 
1) The energies are significantly higher than given by the TF
model with bulk $\xi = 42-44$.  
2) Stabilization of closed shell systems with respect to open shell ones is much weaker
 than in the free gas. However, the density distribution has pronounced fluctuations similar to those of the pure harmonic oscillator density.
3) There is a substantial pairing visible in the odd-even 
binding energy differences, but the magnitude is less than
the bulk pairing parameter associated with a uniform system
of density equal to the central value in the finite system.
 
  We thank A. Bulgac, J. Carlson, M. Forbes, and S. Tan for discussions.
This work was  supported by the U.S. Department of Energy
under Grants DE-FG02-00ER41132 and DE-FC02-07ER41457.  Computations
were performed in part on the NERSC computer facility.


\begin{thebibliography}{}
\expandafter\ifx\csname natexlab\endcsname\relax\def\natexlab#1{#1}\fi
\expandafter\ifx\csname bibnamefont\endcsname\relax
  \def\bibnamefont#1{#1}\fi
\expandafter\ifx\csname bibfnamefont\endcsname\relax
  \def\bibfnamefont#1{#1}\fi
\expandafter\ifx\csname citenamefont\endcsname\relax
  \def\citenamefont#1{#1}\fi
\expandafter\ifx\csname url\endcsname\relax
  \def\url#1{\texttt{#1}}\fi
\expandafter\ifx\csname urlprefix\endcsname\relax\def\urlprefix{URL }\fi
\providecommand{\bibinfo}[2]{#2}
\providecommand{\eprint}[2][]{\url{#2}}

\bibitem{ohara02} K.~M. O'Hara {\it et al.}, Science  {\bf 298}, 2179 (2002).
\bibitem{barten04} M. Bartenstein {\it et al.}, Phys. Rev. Lett.  {\bf 92}, 120401 (2004).
\bibitem{bourdel04} T. Bourdel {\it et al.}, Phys. Rev. Lett. {\bf 93}, 050401 (2004).
\bibitem{chin04} C. Chin {\it et al.}, Science  {\bf 305}, 1128 (2004).
\bibitem{zwier05} M.~W. Zwierlein {\it et al.}, Nature  {\bf 435}, 1047 (2005).
\bibitem{kinast05} J. Kinast {\it et al.}, Science  {\bf 307}, 1296 (2005).
\bibitem{stewart06} J.~T. Stewart, J.~P. Gaebler, C.~A. Regal, and D.~S. Jin,
 Phys. Rev. Lett.  {\bf 97}, 220406 (2006).
\bibitem{ca03} J. Carlson, S.Y. Chang, V.R. Pandharipande, and
K.E. Schmidt, Phys. Rev. Lett. {\bf 91} 050401 (2003).
\bibitem{kalos74} M.~H. Kalos {\it et al}, Phys. Rev. A  {\bf 9} 2178 (1974).
\bibitem{he02a} H. Heiselberg and B. Mottelson, Phys. Rev. Lett.
88 190401 (2002).
\bibitem{he02b} G.M.~Bruun and H. Heiselberg, Phys. Rev. A65 053407
(2002).
\bibitem{we06} F. Werner and Y. Castin, Phys. Rev. Lett. {\bf 97} 150401 (2006).
\bibitem{castin06} F. Werner and Y. Castin, Phys. Rev. A {\bf 74} 053604 (2006).
\bibitem{papen05} T. Papenbrock, Phys. Rev. A {\bf 72} 041603(R) (2005).
\bibitem{br97} M. Brack, R. Bhaduri and R.~K. Bhaduri, ``Semiclassical Physics", (Addison-Wesley, Reading, 1997).
\bibitem{bulgac02a} Y. Yu and A. Bulgac, Phys. Rev. Lett.  {\bf 90},
 222501 (2003).
\bibitem{bulgac02b} A. Bulgac, Phys. Rev. C {\bf 65}, 051305(R) (2002).

\end{thebibliography}
\end{document}